\documentclass[aps,showpacs,showkeys,preprintnumbers,amsmath,epsf,amssymb,epsfig,nofootinbib]{revtex4}
\input epsf.sty
\usepackage{bm}
\usepackage{graphicx}
\draft
\begin{document}

\title{Quintessential inflation from 5D warped product spaces on a dynamical foliation }
\author{$^{1}$ Lucio Fabio P. da Silva \footnote{
E-mail address:luciofabio@fisica.ufpb.br } and
$^1$Jos\'e Edgar Madriz Aguilar\footnote{E-mail address: jemadriz@fisica.ufpb.br}}

\address{$^1$ Departamento de F\'{\i}sica,\\ Universidade
Federal da Para\'{\i}ba. C. P. 5008, CEP 58059-970 Jo\~{a}o Pessoa, Para\'{\i}ba-Brazil.}

\begin{abstract}
Assuming the existence of a 5D purely kinetic scalar field on the class of warped product spaces we investigate the possibility of 
mimic both an inflationary and a quintessential scenarios on 4D hypersurfaces, by implementing a dynamical foliation on the fifth 
coordinate instead of a constant one. We obtain that an induced chaotic inflationary scenario with a geometrically induced scalar 
potential and an induced quasi-vacuum equation of state on 4D dynamical hypersurfaces is possible. While on a constant foliation 
the universe can be considered as matter dominated today, in a family of 4D dynamical hypersurfaces the universe can be passing 
for a period of accelerated expansion with a deceleration parameter nearly -1. This effect of the dynamical foliation results 
negligible at the inflationary epoch allowing for a chaotic inflationary scenario and becomes considerable at the present epoch 
allowing a quintessential scenario.
\end{abstract}


\maketitle
\section{Introduction}

Added interest in theories of gravity in more than four dimensions comes from the fact that matter in 4D has a geometrical origin 
on this scenarios. Among these theoretical settings one can find the induced-matter theory \cite{Wess1,Wess1A,Wess1B} and the 
braneworld scenarios \cite{Maa1,Maa1A}. The basic idea in the induced-matter theory  is that our 4D universe is embedded in a 5D 
space-time in apparent vacuum. The latest condition is understood by the Ricci-flat metric restriction $R_{AB}=0$. In physical 
terms it means that sources of matter in 4D can be geometrically explained by using just one extended extra dimension assuming a 
5D apparent vacuum. Braneworld scenarios can be basically represented by two models. One is the Arkani-Hamed-Dimopoulos-Dvali 
\cite{Dim1,Dim1A}  and the other are the Randall-Sundrum braneworld models I and II \cite{RS12,RS12A}. In the context of 
braneworld models our universe is embedded in a higher dimensional space-time called the bulk while our observable universe lies 
on a 4D brane hypersurface.  The main motivation of the Randall-Sundrum model I is to solve the gauge hierarchy problem in a 
geometrical manner instead of using symmetry principles \cite{Maa1,Maa1A}. Metrics in Randall-Sundrum models and in general in 
braneworld scenarios are constructed employing the class of warped product metrics. This fact makes important the study of this 
kind of metrics even in other theories of gravitation.\\

Matter confinement in braneworld cosmologies yields a cosmological evolution in the brane which is different from the typical FRW 
evolution generating problems for instance with the usual nucleosynthesis scenario \cite{Lan}. In an attempt of finding a solution 
to this problem a negative cosmological constant in the bulk was introduced generating a fine tuning between the brane tension and 
the bulk cosmological constant. This fine-tuning is an undesirable characteristic of many braneworld models. This kind of problems 
suggested the study of cosmological braneworld models with dynamical fields in the bulk. Among them we can find models with bulk 
scalar fields.\\

As it was shown in  \cite{Mio1} when we consider a 5D purely kinetic scalar field on warped product spaces, it is possible to 
recover a standard inflationary scenario on 4D hypersurfaces given by establishing a constant foliation on the fifth coordinate, 
where the scalar potential of the inflaton field is geometrically induced and  the equation of state of the induced matter is 
given by $P_{IM}=-\rho _{IM}$. There quantum confinement of the inflaton modes is achieved naturally for at least a class of 
warping factors. However there is no possibility to apply that formalism to describe in the same model the inflationary phase and 
the present period of accelerated expansion of the universe known as quintessential expansion . \\

A great quantity of cosmological data accumulated during the last years are supporting the idea that nowadays the universe is 
passing for a phase of accelerated expansion. The Hubble diagram of Type Ia Supernovae (SNeIa), measured by both the Supernovae 
Cosmology Project and the High-z Team up to redshift $z\sim 1$, was the first evidence that the expansion of the universe is 
currently accelerated \cite{Salvatore}. The simplest explanation of this phenomena is the cosmological constant $\Lambda$. However 
in the $\Lambda$CDM cosmological model it is not clear why the inferred value of $\Lambda$ is 120 orders of magnitude lower with 
respect to the value inferred in particle physics.  As another tentative solution many authors have proposed to replace the 
cosmological constant with a scalar field rolling down its potential generating a model known as quintessence \cite{Pad}. \\

In this letter we are interested in investigate the possibility to extend  the formalism developed in \cite{Mio1} in order to 
study  quintessential and inflationary scenarios in a more geometrical manner. To achieve this we shall assume that the 5D 
space-time can be dynamically foliated by a family of 4D dynamical hypersurfaces $\Sigma:\psi =f(x^{0})$, being $\psi$ the 
space-like fifth coordinate and $x^{0}$ a time-like coordinate. The possibility of having dynamical hypersurfaces has been 
investigated by J. Ponce de Leon in \cite{JPLeon,JPLeonA}. In this letter we shall follow the Ponce de Leon's mechanism of 
implementing dynamical foliations not just for 5D Ricci-flat metrics but for 5D warped product metrics. We have organized this 
letter as follows. Section I is a brief introduction. Section II is devoted to a general formalism describing how to implement 
dynamical foliations and induce effective scalar potentials geometrically on the class of warped product metrics. In section III 
we approach a cosmological scenario from a particular warped geometry. In subsections A and B, inflationary and quintessential 
scenarios are respectively obtained. Finally, section IV is reserved to some final comments and conclusions. In our conventions 
capital Latin indices run from 0 to 4, small Latin indices from 1 to 3 and Greek indices from 0 to 3. Also we adopt the metric 
signature $(+,-,-,-,-)$.

\section{General formalism}

We consider a 5D space-time with metric $g_{AB}$, endowed with a purely kinetic scalar field $\varphi$ minimally coupled to 
gravity. Thus, in the local coordinates  $y^{A}=(x^{\alpha},\psi)$  the 5D-action reads
\begin{equation}\label{q1}
^{(5)}{\cal S}=\int d^{\,4}x \,d\psi \, \sqrt{^{(5)}g}\left[\frac{^{(5)}R}{2\kappa^{2}}-\frac{1}{2}g^{AB}\varphi _{,A}\varphi 
_{,B}\right],
\end{equation}
where $^{(5)}g$ is the determinant of the metric $g_{AB}$, $^{(5)}R$ is the scalar curvature and $\kappa$ being the 5D 
gravitational coupling. The field equations derived from the action (\ref{q1}) are of the form
\begin{eqnarray}\label{q2}
&&G_{AB}=\kappa ^{2}T_{AB},\\
\label{n1}
&&^{(5)}\Box \varphi\equiv \frac{1}{\sqrt{^{(5)}g}}\frac{\partial}{\partial x^{A}}\left[\sqrt{^{(5)}g}g^{AB}\varphi 
_{,B}\right]=0.
\end{eqnarray}
being $T_{AB}^{(\varphi)}=\varphi _{,A}\varphi _{,B}-\frac{1}{2}g_{AB}\varphi ^{,C}\varphi _{,C}$ the energy momentum tensor for 
the purely kinetic scalar field $\varphi$ and $G_{AB}=R_{AB}-(1/2)g_{AB}\,^{(5)}R$ the 5D Einstein tensor.\\

Now let us to consider the class of warped geometries which are given by the line element
\begin{equation}\label{q5}
dS^{2}=e^{2A(\psi)}h_{\alpha\beta}dx^{\alpha}dx^{\beta}-d\psi^{2},
\end{equation}
where the real value function $A(\psi)$ is a warping factor and the 4D-part $h_{\alpha\beta}$ of the 5D metric $g_{AB}$ does not 
depend of the fifth coordinate, i.e. $h_{\alpha\beta}=h_{\alpha\beta}(x)$. Clearly, in (\ref{q5}) the fifth coordinate $\psi$ has 
spatial units and $A(\psi)$ is dimensionless.\\

On this metric background equation (\ref{n1}) becomes
\begin{equation}\label{n2}
^{(4)}\Box \varphi -e^{-2A(\psi)}\frac{\partial}{\partial \psi}\left[e^{4A(\psi)}\frac{\partial \varphi}{\partial \psi}\right]=0,
\end{equation}
where $^{(4)}\Box \varphi =(1/\sqrt{-h})(\partial /\partial x^{\mu})(\sqrt{-h}h^{\mu\nu})\varphi _{,\nu}$. Assuming separability 
of the scalar field $\varphi (x,\psi)=\Gamma(\psi)\phi(x)$, the dynamics of the field on the fifth dimension is determined by 
\cite{Mio1}
\begin{equation}\label{n3}
\overset{\star\star}{\Gamma}+4\overset{\star}{A}(\psi)\overset{\star}{\Gamma}-\alpha e^{-2A(\psi)}\Gamma=0,
\end{equation}
with $\alpha$ a separation constant and the star $(\star)$ denoting derivative with respect to the fifth dimension. This is a 
second order differential equation, which depending of the warping factor chosen in principle can be solved. Therefore, in general  
the 5D dynamics of $\varphi$ depends strongly of the class of warping factors considered.\\

In order to obtain the 4D effective dynamics, we consider a special family of 4D-hypersurfaces in the same manner as they were 
introduced by J. Ponce de Leon in \cite{JPLeon,JPLeonA}, called dynamical foliations. This way, assuming that the 5D space-time 
can be foliated by a family of dynamical hypersurfaces $\Sigma :\psi=\psi(x^{0})$, being $x^{0}$ a time-like local coordinate, the 
induced metric on $\Sigma$ reads
\begin{equation}\label{q6}
ds_{\Sigma}^{2}=\left[e^{2A[\psi(x^0)]}h_{00}(x)-\left(\frac{d\psi}{dx^0}\right)^{2}\right](dx^0)^{2}+e^{2A[\psi(x^{0})]}h_{ij}(x)
dx^{i}dx^{j}.
\end{equation}
Thus, in order to recover an effective 4D line element $ds^{2}=\gamma _{\mu\nu}(x)dx^{\mu}dx^{\nu}$ on the hypersurface $\Sigma$, 
whose explicit form depends of the problem addressed, the continuity of $g_{AB}$ across $\Sigma $ requires \cite{JPLeon,JPLeonA}
\begin{eqnarray}\label{q7}
e^{2A[\psi(x^0)]}h_{00}(x)-\left(\frac{d\psi}{dx^0}\right)^{2}&=&\gamma _{00}(x),\\
\label{q8}
e^{2A[\psi(x^0)]}h_{ij}(x)&=&\gamma _{ij}(x).
\end{eqnarray}
Solutions of (\ref{q7}) determine the dynamical foliations $\psi=\psi(x^{0})$ allowed by the 5D geometry, while equation 
(\ref{q8}) describes the contribution of every dynamical foliation $\psi=\psi(x^0)$ to the 3D spatial part of the effective 
metric, $\gamma _{ij}(x)$.\\

On a generic hypersurface $\Sigma :\psi=\psi (x^{0})$, the effective 4D action reads
\begin{equation}\label{n4}
^{(4)}{\cal S}_{eff}=\int d^{4}x \sqrt{-\,^{ (\Sigma)}\!g}\left[\,\frac{^{(4)}\!R_{\Sigma}}{16\pi G}+{\cal L}_{IM}+\,^{(e)}\!{\cal 
L}_{IM}+\frac{1}{2}\,^{(\Sigma)}\!g^{\alpha\beta}\varphi _{eff\, ,\alpha}\varphi _{eff\, ,\beta}-V(\varphi _{eff})\right],
\end{equation}
where $^{(\Sigma)}g_{\alpha\beta}$ is the tensor metric on $\Sigma$, ${\cal L}_{IM}$ is a typical induced matter lagrangian 
density, $^{(e)}{\cal L}_{IM}$ is a new contribution to ${\cal L}_{IM}$ due to the dynamical foliation considered $\psi =f(x^0)$,  
$\varphi _{eff}[x^{0},x^{i},f(x^{0})]=\left.\varphi(x,\psi)\right|_{\psi=f(x^0)}$ is the effective scalar field induced on 
$\Sigma$ and the effective 4D induced scalar potential is given by
\begin{equation}\label{n5}
V(\varphi 
_{eff})=\left.\frac{1}{2}\left(\frac{\partial\varphi}{\partial\psi}\right)^{2}\right|_{\psi=f(x^{0})}=\frac{1}{2}\,m^{2}(t)\, 
\varphi _{eff}^{2},
\end{equation}
being $m^{2}(t)=[\overset{\star}{\Gamma}(\psi)/\Gamma(\psi) ]^{2}|_{\psi =f(t)}$ a dynamical mass associated to $\varphi _{eff}$ 
which is geometrical in origin.

\section{A cosmological scenario}

Since we are interested in cosmological solutions, we will assume that the metric $h_{\alpha\beta}$ is spatially 3D homogeneous 
and isotropic. Choosing for simplicity $h_{\alpha\beta}$ as a spatially flat FRW metric, the 5D line element (\ref{q5}) becomes
\begin{equation}\label{q6}
dS^{2}=e^{2A(\psi)}\left[dt^{2}-a^{2}(t)dr^{2}\right]-d\psi^{2},
\end{equation}
where $a(t)$ is the scale factor and $dr^{2}=dX^{2}+dY^{2}+dZ^{2}$, being $[X,Y,Z]$  cartesian coordinates. On a generic dynamical 
hypersurface $\Sigma _{c}: \psi=f(t)$ the induced 4D metric $^{(\Sigma _c)}g_{\alpha\beta}$ is given by the line element
\begin{equation}\label{cos1}
ds_{\Sigma _c}^{2}=\left[e^{2A[f(t)]}-\dot{f}^{2}(t)\right]dt^{2}-e^{2A[f(t)]}a^{2}(t)dr^{2},
\end{equation}
with the dot denoting partial derivative with respect to the cosmic time $t$. In order to obtain a spatially flat FRW on the 4D 
hypersurface $\Sigma _{c}$, where the line element is
\begin{equation}\label{cos2}
ds_{\Sigma _c}^{2}=dt^{2}-b^{2}(t)dr^{2},
\end{equation}
the continuity conditions across $\Sigma _{c}$ (\ref{q7}) and (\ref{q8}) read
\begin{eqnarray}\label{cos3}
e^{2A[f(t)]}-\dot{f}^{2}(t)&=&1,\\
\label{cos4}
e^{2A[f(t)]}a^{2}(t)&=& b^{2}(t),
\end{eqnarray}
being $b(t)$ an affective scale factor. In this case, we can see from equation (\ref{cos4}) that the typical scale factor $a(t)$ 
is dynamically rescaled as a consequence of taking a dynamical foliation. \\

From the action (\ref{n4}) the  effective 4D Friedmann equations on $\Sigma _{c}$ are
\begin{eqnarray}\label{n6}
3\left(\frac{\dot{b}}{b}\right)^{2}&=&8\pi G \left[\rho _{IM}+\rho _{e}+\rho _{s}\right],\\
\label{n7}
2\frac{\ddot{b}}{b}+\left(\frac{\dot{b}}{b}\right)^{2}&=&-8\pi G\left[P_{IM}+P_{e}+P_{s}\right],
\end{eqnarray}
where the induced matter density and pressure $\rho _{IM}$ and $P_{IM}$, the extra density and pressure terms coming due to the 
dynamical foliation $\rho _{e}$ and $P_{e}$, and the density and pressure of the effective scalar field $\rho _{s}$ and $P_{s}$, 
are given by
\begin{eqnarray}\label{n8}
\rho _{IM}&=&-P_{IM}=\frac{3}{8\pi G}(A''+2A'^{2})\\
\label{n9}
\rho _{e}&=&\frac{3}{8\pi G}\left[2\frac{\dot{b}}{b}A'\dot{f}+(A'\dot{f})^{2}\right]\\
\label{n10}
P_{e}&=& -\frac{1}{8\pi G}\left[6A'\frac{\dot{b}}{b}\dot{f}+(2A''+3A'^{2})\dot{f}^{2}+2A'\ddot{f}\right]\\
\label{n12}
\rho _{s}&=&\frac{1}{2}\,\left(\dot{\varphi}_{eff}+\varphi '_{eff}\dot{f}\right)^{2}+\frac{e^{2A(f)}}{2b^{2}}\left(\nabla 
_{r}\varphi _{eff}\right)^{2}+V(\varphi _{eff}),\\
\label{n11}
P _{s}&=&\frac{1}{2}\,\left(\dot{\varphi}_{eff}+\varphi '_{eff}\dot{f}\right)^{2}+\frac{e^{2A(f)}}{2b^{2}}\left(\nabla _{r}\varphi 
_{eff}\right)^{2}-V(\varphi _{eff})\\
\end{eqnarray}
where the prime $(')$ denotes $(\partial /\partial \psi)$ evaluated at $\psi =f(t)$. Besides, the field equation for the effective 
4D scalar field reads
\begin{equation}\label{n13}
\ddot{\varphi}_{eff}+3\frac{\dot{b}}{b}\dot{\varphi}_{eff}-e^{2A(f)}b^{-2}\nabla _{r}^{2}\varphi _{eff}+\frac{dV(\varphi 
_{eff})}{d\varphi _{eff}}+\left[2\dot{\varphi}'_{eff}+3\frac{\dot{b}}{b}\varphi' 
_{eff}-3A'\dot{\varphi}_{eff}\right]\dot{f}+\left(\varphi''_{eff}-3A'\varphi'_{eff}\right)\dot{f}^{2}+2\varphi'_{eff}\ddot{f}=0.
\end{equation}
Hence, the 4D effective cosmological dynamics is described by equations (\ref{n6}), (\ref{n7}) and (\ref{n13}). Clearly, when we 
consider a constant foliation $\psi =\psi _{0}$ instead of a dynamical one, equations (\ref{n6}) and (\ref{n7}) reduce to the ones 
obtained in \cite{Mio1}, while equation (\ref{n13}) reduces to a typical scalar field equation for an effective scalar field 
$\varphi _{eff}$.

\subsection{4D effective inflation}

In order to obtain inflationary solutions on the scenario previously described, we consider a homogeneous effective scalar field, 
$(1/2)(\nabla _{r}\varphi _{eff})^{2}\ll V(\varphi _{eff})$. In addition we assume a type of slow-roll conditions 
$(1/2)(\dot{\varphi}_{eff}+\varphi '_{eff}\dot{f})^{2} \ll V(\varphi _{eff})$. Thus, defining the total energy density $\rho _{T}$ 
and pressure $P_{T}$ respectively as $\rho _{T}=\rho _{IM}+\rho _{e}+\rho _{s}$ and $P_{T}=P_{IM}+P_{e}+P_{s}$, the general 
equation of state can be written as $P_{T}=\omega (t)\rho _{T}$ where
\begin{equation}\label{n14}
\omega (t)\simeq-\left[1+ \frac{2A''\dot{f}^{2}+2A'\ddot{f}}{3A''+6A'^{2}+6A'(\dot{b}/b)\dot{f}+(A'\dot{f})^{2}+M_{Pl}^{-2}V( 
\varphi _{eff})}\right]
\end{equation}
being $M_{Pl}=(8\pi G)^{-1/2}$ the reduced Planckian mass. As the dynamical foliation $\psi =f(t)$, according to (\ref{cos3}),  
depends directly of the warping factor $A(\psi)|_{\psi =f(t)}$, we can say that in principle for every warping factor  we have a 
different type of induced matter which satisfy the equation of state (\ref{n14}).\\

Now, in order to obtain a particular cosmological scenario let us to consider the warping factor $A(\psi)=ln (\psi /\psi _{0})$, 
with $\psi _{0}$ an arbitrary dimensionalization constant with units of lenght. Thus, the 5D line element (\ref{q6}) becomes
\begin{equation}\label{n15}
ds^{2}=\left(\frac{\psi}{\psi _{0}}\right)^{2}[dt^{2}-a^{2}(t)dr^{2}]-d\psi ^{2}.
\end{equation}
Solving equation (\ref{cos3}) for this $A(\psi)$ we obtain the solutions
\begin{equation}\label{n16}
f(t)=\pm \psi _{0},\qquad f(t)=\frac{\psi _0}{2}\left[1+e^{-2\Delta t/\psi _0}\right]e^{\Delta t/\psi _0},
\end{equation}
where $\Delta t=t-t_{0}$, being $t_{0}$ the time when inflation begins. The first two solutions correspond to constant foliations 
and the last corresponds to a dynamical one. On the other hand, as it is well-known the inflationary period is pretty short ( 
$\sim 10 ^{-35}\,seg$ ) \cite{Guth1}, this way during this epoch $\Delta t <<1$, thus the dynamical foliation in (\ref{n16}) can 
be approximated as $f(t)\simeq\psi _0[1-\left(\Delta t/\psi _0\right)^{2}]\simeq \psi _{0}$, which means that practically during 
inflation this dynamical foliation behaves as a constant foliation. This way according to (\ref{n14}) we can recover a 
quasi-vacuum equation of state $P_{T}\simeq -\rho _{T}$ during inflation. \\

With this warping factor, the equation (\ref{n3}) has the general solution
\begin{equation}\label{n17}
\Gamma (\psi)=C_{1}\psi ^{\sigma}+C_{2}\psi ^{-\sigma},
\end{equation}
being $C_{1}$ and $C_{2}$ integration constants and $\sigma =(1/2)[-3+\sqrt{9-4\alpha\psi _{0}^{2}}\,]$. Choosing $C_{2}=0$ and 
using (\ref{n5}) the 4D induced potential reads
\begin{equation}\label{n18}
V(\varphi _{eff})=\frac{1}{2}\,m_{0}^{2}\,\varphi _{eff}^{2},
\end{equation}
where $m_{0}=\sigma\psi _{0}^{-1}$ is a constant geometrical mass associated to the effective inflaton field $\varphi _{eff}$. 
Thus, the dynamical equation (\ref{n13}) reduces to
\begin{equation}\label{n19}
\ddot{\varphi }_{eff}+3H(t)\dot{\varphi}_{eff}-\frac{1}{a^2}\nabla _{r}^{2}\varphi _{eff}+m_{0}^{2}\varphi _{eff}= 0.
\end{equation}
This equation describes the typical dynamics on a chaotic inflationary scenario.
Chaotic inflation is a well-studied scenario and its dynamics can be found for example in 
\cite{Linde1,Linde1A,Linde1B,Linde1C,Linde1D}. In our case the inflaton mass $m_{0}$ can have a realistic value. For $\alpha 
\simeq -0.9\times 10^{-11}\,yrs^{-2}$ and $\psi _{0}=0.5\times 10^{10}\,yrs=0.584\times 10^{60}M_{Pl}^{-1}$  the inflaton mass has 
the value $m_{0}\simeq 3\times 10 ^{-6}\, M_{Pl}$, which is a realistic value according to \cite{Linde2}.

\subsection{ A quintessential scenario}

\begin{figure}
\includegraphics[width=12cm]{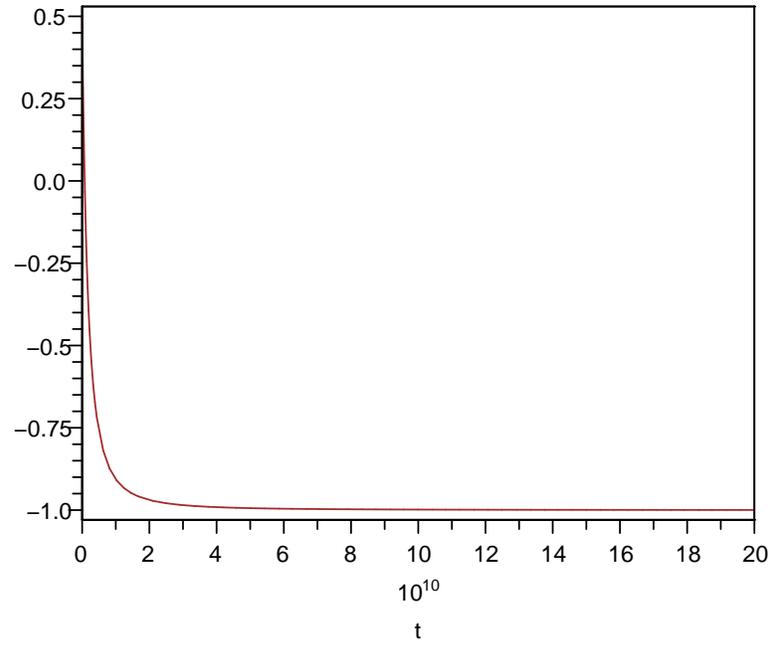}
\caption{\label{fig1} The figure shows the behavior of the deceleration parameter $q(t)$ with time $t$ during the present epoch, 
where we have taken $\psi _0 =0.5\times 10^{10}\, yrs$, $p=2/3$ and t is given in years.}
\end{figure}

One of the current problems in modern cosmology is to explain the present period of accelerated expansion of the universe 
supported by numerous observations \cite{darke,darkeA,darkeB,darkeC,darkeD,darkeE,darkeF}. A quintessential scenario can be 
obtained from the present formalism, where the quintessential accelerated expansion depends of the warping factor and is due to 
the dynamics of the fifth dimension i.e. it is due to we are considering a dynamical foliation of the warped 5D space-time.\\
The deceleration parameter $q$ using (\ref{cos4}) reads
\begin{equation}\label{n20}
q=-\frac{\ddot{b}b}{\dot{b}^{2}}=-\frac{a\ddot{a}+a\left(\dot{a}A'+A''\dot{f}+A'^{2}\dot{f}\right)\dot{f}+aA'\ddot{f}}{\left( 
\dot{a}+A'\dot{f}\right) ^{2}}.
\end{equation}
At the present epoch $\Delta t \gg 1$, therefore the dynamical foliation (\ref{n16}) can be approximated by $f(t)\simeq (\psi 
_{0}/2)e^{\Delta t/\psi _{0}}$. Thus, considering $A(\psi)=ln(\psi /\psi _0)$ and a power-law for $a(t)\simeq t^{p}$,  the 
deceleration parameter (\ref{n20}) becomes
\begin{equation}\label{n21}
q=-\left[1-\frac{p\psi _{0}^{2}}{t^{2}+2p\psi _{0}t+p^{2}\psi _{0}^{2}}\right].
\end{equation}
According to (\ref{n20}) and (\ref{n21}), the new contribution comes from the dynamical foliation $\psi =f(t)$. Thus we can say 
that a universe which is currently matter dominated $(p=2/3)$ on a constant foliation, can be now in a quintessential period on a 
dynamical one.  Hence, for  $H_{0}= 0.7\times 10^{-10}yr^{-1}$, $t_{0}= 1.4\times 10^{10}\,yrs$ and $\psi _{0}=0.5\times 10^{10}$, 
the deceleration parameter (\ref{n21}) at the present epoch would be $q_{0}=-0.9445$. The behavior  of $q$ as a function of $t$ in 
the present epoch is shown in the figure [\ref{fig1}]. We can see that $q$ tends asymptotically to $-1$ with time.\\
Inserting (\ref{n20}) in  $\omega(t)=[2q(t)-1]/3$ we obtain
\begin{equation}\label{n22}
\omega (t)= -1+\frac{2p\psi _{0}^{2}}{3[t_{0}^{2}+2p\psi _{0}t_{0}+p^{2}\psi _{0}^{2}]},
\end{equation}
which according to the previous data has the present value $\omega _{0}=-0.963$. These results fit into the current observational 
data according to \cite{data,dataA,dataB,dataC}

\section{Final comments}

In this letter we have investigated the possibility of mimic both the inflationary and the present quintessence epochs from a 5D 
warped product space equipped with a purely kinetic scalar field, by implementing a dynamical foliation instead of a constant one. 
We have obtained that if we consider a matter-dominated universe at the present time on the constant foliation $\Sigma _{0}:\psi 
=\psi _0$, it is possible to obtain a quintessential scenario where the accelerated expansion is geometrically generated by the 
dynamical foliation $\Sigma: \psi=f(t)$. The explicit form of the function $f(t)$ is given by the continuity conditions introduced 
by J. Ponce de Leon in \cite{JPLeon,JPLeonA}. For a warping factor of the form $A(\psi)=ln(\psi/\psi _0)$ the dynamical foliation 
behaves as a constant one for early times and in particular during the inflationary phase. However it becomes completely dynamical 
at the present epoch generating geometrically a quintessential expansion with $q\simeq -1$. The 4D induced inflationary stage 
results a scenario of chaotic inflation with a geometrically induced scalar potential $V(\varphi _{eff})=(1/2)m_{0}^{2}\varphi 
_{eff}^{2}$, where a realistic value of $m_{0}\simeq 3\times 10 ^{-6}\,M_{Pl}$ can be achieved. Finally we show a plot with the 
behavior  of the deceleration parameter $q$ during the quintessential period.

\begin{acknowledgments}
 L. F. P. S. and J. E. M. A. acknowledge CNPq-CLAF and UFPB
for financial support.
\end{acknowledgments}

\end{document}